\documentclass[epj]{svjour}
\usepackage{graphics}
\begin{document}
\title{On the background in the $\gamma p\to \omega(\pi^0\gamma) p$ 
reaction \\ and mixed event simulation}
\author{M. Kaskulov 
\inst{1} 
\and E. Hern\'andez \inst{2} 
\and E. Oset \inst{3}
}                     
%
%
\institute{Institut f\"ur Theoretische Physik, Universit\"at Giessen,
             D-35392 Giessen, Germany \and Grupo de F\'\i sica Nuclear, Departamento de F\'\i sica
Fundamental e IUFFyM, Universidad de Salamanca, E-37008 Salamanca,
Spain \and Departamento de F\'\i sica Te\'orica e
IFIC, Centro Mixto Universidad de Valencia-CSIC, Institutos de Investigaci\'on
de Paterna, Aptd. 22085, E-46071 Valencia, Spain}
\date{Received: / Revised version: }
%
\abstract{
In this paper we evaluate sources of background for the $\gamma p \to \omega p$,
with the $\omega$ detected through its $\pi^0 \gamma$ decay channel, to compare with
the experiment carried out at ELSA. We find background from 
$\gamma p \to \pi^0 \pi^0 p$ followed by decay of a $\pi^0$ into two $\gamma$,
recombining one $\pi^0$ and one $\gamma$, and from the 
$\gamma p \to \pi^0 \eta p$ reaction with subsequent decay of the $\eta$ into
two photons. This background accounts for the data at $\pi^0 \gamma$ invariant
masses beyond 700 MeV, but strength is missing at lower invariant masses which
was attributed to photon misidentification events, which we simulate to get a
good reproduction of the experimental background. Once this is done, we perform
an event mixing simulation to reproduce the calculated background and we find
that the method provides a good description of the background at low 
$\pi^0 \gamma$ invariant masses but fakes the background at high invariant
masses,
making background events at low invariant masses, which are due to $\gamma$
misidentification events, responsible for the background at high invariant
masses which is due to the  $\gamma p \to \pi^0 \pi^0 p$ and 
$\gamma p \to \pi^0 \eta p$ reactions.
\PACS{
      {13.60.Le}{describing text of that key}   \and
      {25.20.Lj}{describing text of that key}
     } 
} 
\maketitle
\section{Introduction}
\label{intro}
The interaction of vector mesons with nuclear matter has attracted attention for
a long time and has been tied to fundamental aspects of QCD. Yet, the
theoretical models offer a large variety of results from a large attraction to a
large repulsion. Early results on this issue within the Nambu Jona Lasinio
model produced no shift of the masses \cite{Bernard:1988db}  while, using
qualitative arguments, a universal large shift of the mass was suggested in 
\cite{Brown:1991kk}. More recent detailed calculations show no shift of the mass
of the $\rho$ meson in matter \cite{Rapp:1997fs,Urban:1999im,Cabrera:2000dx}.
Experimentally the situation has undergone big steps recently with the NA60
collaboration reporting a null shift of the $\rho$ mass in the medium
\cite{Arnaldi:2006jq,Damjanovic:2007qm} in the dilepton spectra of heavy ion
reactions and also a null shift in the $\gamma$ induced dilepton production at
CLASS \cite{Wood:2008ee}. On the other hand the KEK team had earlier reported 
 an attractive mass shift of the $\rho$ in \cite{Muto:2005za,Naruki:2005kd}. 
As explained in detail in \cite{Wood:2008ee}, the different conclusions can be
traced back to the way the background is subtracted. Thus, the treatment of the
background is an essential part of the investigation of the vector meson
properties in nuclei.  
  The case of the $\omega$ in the medium is more obscure. Theoretically there
are about twenty different works with claims from large attraction to large
repulsion (see \cite{Muhlich:2003tj,Kaskulov:2006zc,Kaskulov:2006fi} for details). 
Experimentally there are claims of a large shift of the mass of the $\omega$
from the study of the photon induced $\omega$ production in nuclei, with the
$\omega$ detected through its $\pi^0 \gamma$ decay channel \cite{david,prl}.
However, it was shown in \cite{Kaskulov:2006zc} that the shift could just be a consequence
of a particular choice of background subtraction and that other reasonable 
choices led to different conclusions. For instance, choosing a background
in the nuclear case proportional to the background on the proton in the region
below the $\omega$ peak, the experimental data could be explained without a 
shift of the $\omega$ mass in the medium. 

The method to determine the mass shift is
very different from the one used to determine the width in the medium. This
latter one relies upon the production cross section in different nuclei, which
leads to the transparency ratio that allows to determine a large width of the
$\omega$ in the medium \cite{Kaskulov:2006zc,:2008xy}. By contrast the
measurement of the mass requires the analysis of the shape of the 
invariant mass distribution,
which is barely affected in nuclei because practically all $\omega$ 
decays occur in the nuclear surface or outside the nucleus. 

    The discussion on the issue of the $\omega$ mass shift was followed
     by the evaluation of the
background with the mixed event technique in \cite{metagerice}. There the
background for a nuclear target was evaluated and found to be the same as
assumed in \cite{prl}, and again it was concluded that
the data demanded a shift of the $\omega$ mass in the medium. 

  The former discussion indicates that the treatment of background is an
essential issue in this problem. In view of this we decided to face the problem
and investigate the details on how the mixed event technique works in the
present case. For this we followed the strategy of evaluating the background for
the proton target. We could trace two sources of background that account for the
experimental cross section at $\pi^0 \gamma$ invariant masses of the order of
the  $\omega$ mass 
and beyond. The rest of the background at lower invariant masses was simulated
to account for $\gamma$ misidentification events, as found in \cite{david}. Once
a background consistent with the experiment is obtained theoretically,  we
apply the mixed event technique to obtain the background and compare it with the
theoretical one. In this way
we can determine the ability of the mixed event method to reproduce the
background in this reaction. The results that we find are that the method can
provide the background at low invariant masses, where the cross sections are
large, but it actually fakes the background in the region of invariant masses 
around and beyond the $\omega$ mass. We show that the mixed event generated 
background in that region is completely tied to the real events at low 
$\pi^0\gamma$ invariant masses where the $d\sigma/dM_{\pi^0\gamma}$ distribution
is larger. 
As a result, we show that the distribution obtained with the mixed event method
 in the region of invariant masses around and
beyond the $\omega$ mass is largely insensitive to the actual background 
contributing in that region. But this is 
 precisely 
the region where the background is needed to
determine changes of the $\omega$ signal in the nuclear medium. As a consequence
we clearly show that the mixed event technique is unsuited in the present case
as an instrument to determine possible  shifts of the  $\omega$ mass 
in the medium.  
 
In the meantime, a  recent reanalysis of the
background of the reaction of \cite{david,prl} done in \cite{marianahad} 
concludes, however, that one cannot claim a shift of the $\omega$ mass from 
this experiment.

\section{Background sources in the reaction
$\gamma p\to \omega p\ (\omega\to \pi^0\gamma)$}
The reaction that we study is $\gamma p\to \omega p$ where the 
$\omega$ is detected through its $\omega\to \pi^0\gamma$ decay mode.
This is the reaction studied in the CBELSA/TAPS experiment. According to the study
in \cite{david}, in the region of the reconstructed invariant mass of the 
$\pi^0\gamma$ one of the main sources of background comes from the 
$\gamma p\to \pi^0\pi^0 p$ reaction followed by the decay of any of the two 
$\pi^0$ into $\gamma\gamma$. Then, background events appear from the combination
of one of these photons and the remaining $\pi^0$. Another source is the 
$\gamma p\to \pi^0\eta p$ reaction. We evaluate the cross section for those two
processes in the following subsection.
\subsection{The $\gamma p\to \pi^0\pi^0 p$ reaction}
This reaction has been thoroughly studied  at
Mainz~\cite{Braghieri:1994rf,Harter:1997jq,mainz,Schadmand:2005ji} and more
recently at ElSA~\cite{elsa,thoma}, GRAAL \cite{Ajaka:2007zz} and Jefferson 
Lab~\cite{Ripani:2000aw,jeff,Burkert:2007kn}. 
We are interested
not only in the cross section for the reaction but at the same time to have an
event generator that provides events weighed by their probability determined by
the available phase space. For this purpose the Monte Carlo evaluation of the
cross section integral is the most suited algorithm since it provides the events
allowed by phase space properly weighted and the cross section in the end.

The $\gamma p\to \pi^0\pi^0 p$ cross section is given by
\begin{eqnarray}
\sigma =\frac{M^2_N}{s-M^2_N}\int\frac{d^3p_1}{(2\pi)^3}\int\frac{d^3p_2}{(2\pi)^3}
\int\frac{d^3p_3}{(2\pi)^3} \nonumber \\
\times \frac{1}{2E( p_1)}\frac{1}{2E( p_2)}
\frac{1}{2E( p_3)}\left|T\right|^2  \nonumber \\
\times
(2\pi)^4 \delta(p_\gamma+p_p-p_1-p_2-p_3),
\label{eq:cros_sec}
\end{eqnarray}
 which includes the 1/2 symmetry factor to account for the two identical pions in the final state.
 The variables $p_1,\,p_2,\,p_3$ are the momenta of the final proton and
 the two  $\pi^0$ respectively and $T$ stands for the transition matrix for
 the $\gamma p\to \pi^0\pi^0 p$  process. In the  $\left|T\right|^2$ factor 
 a sum over final spins and a proper average over initial spins
 is implicit.
 
 Since $ d^3p/E( p)$ is a Lorentz invariant measure, we proceed to evaluate
 the last  two integrals in Eq.~(\ref{eq:cros_sec}) in the reference frame where
 $\vec p_\gamma+\vec p_p-\vec p_1=\vec 0$ which guarantees that 
 $\vec p_2+\vec p_3=\vec 0$. The cross section is then written as
\begin{eqnarray}
\sigma=\frac{M^2_N}{s-M^2_N}\int\frac{d^3p_1}{(2\pi)^3}\frac{1}{2E( p_1)}
\theta(M_{23}-2m_{\pi^0}) \nonumber \\
\times \int d\Omega\ \tilde p_2 
\frac{1}{16\pi^2}\frac{1}{M_{23}}\left|T\right|^2,
\label{eq:cros_sec_2}
\end{eqnarray}
where $M_{23}$  is the invariant mass of the two $\pi^0$ given by
\begin{eqnarray}
M_{23}^2=(p_\gamma+p_p-p_1)^2=s+M^2_N-2(p_\gamma+p_p)p_1.
\label{eq:invmass}
\end{eqnarray}
We shall work in the laboratory (lab) frame that allows us to implement easily all
the experimental cuts. The variable $\tilde p_2$ in Eq.~(\ref{eq:cros_sec_2})
is the $\pi^0$ momentum in the $\pi^0\pi^0$ rest frame
\begin{eqnarray}
\tilde p_2=\frac{\lambda^{\frac12}(M_{23}^2,m_{\pi^0}^2,m_{\pi^0}^2)}{2M_{23}},
\end{eqnarray}
and $d\Omega$ is performed in the $\pi^0\pi^0$ rest frame.

The $\gamma p\to \pi^0\pi^0 p$ reaction has been studied theoretically at
$E_\gamma< 0.8$\,GeV~\cite{tejedor,Bernard:1996ft,roca,nacher,Ochi:1997ev} 
and at higher energies
$E_\gamma>1.5$\,GeV in~\cite{Ripani:2000aw,Roberts:2004mn}. We do not need any of these sophisticated
models here. Our final goal is to see how much background comes form this source
and to have an event generator by means of which we can study how the mixed
event technique works in the present case. For this purpose it is enough to
consider $|T|^2$ to be a constant over the phase space and fit its value to
reproduce the experimental results for the cross section in the region of
interest to us. We take the cross section for $\gamma p\to \pi^0\pi^0 p$
at the needed photon energies from~\cite{elsa,thoma}. 

The next step is to write  $\tilde p_2$ in the lab frame. We have in 
the $\pi^0\pi^0$ rest frame
\begin{eqnarray}
\vec{ \tilde p}_2 &=& \tilde p_2\left\{\begin{array}{c}
sin\theta\,cos\varphi\\
sin\theta\,sin\varphi\\
cos\theta
\end{array}\right\} \\ \vec{\tilde p}_3 &=&-\vec{\tilde p}_2
\end{eqnarray}
with $\theta,\varphi$ angles in the  $\pi^0\pi^0$ rest frame. We
then perform a boost of $\vec{\tilde p_2}$ to the lab frame
where $\vec p_2+\vec p_3=\vec p_\gamma+\vec p_p-\vec p_1=\vec P$
\begin{eqnarray}
\vec p_2=\left[\left(\frac{E_{23}}{M_{23}}-1\right)\frac{\vec{\tilde
p}_2\cdot\vec P}{\vec P^2}+\frac{\tilde p_2^0}{M_{23}}
\right]\vec P+\vec{\tilde p}_2,
\end{eqnarray}
where $E_{23}$ is the two pion energy in the lab frame
$E_{23}=(M_{23}^2+\vec P^2)^{\frac12}$. Similarly we boost
$\vec{\tilde p}_3$ to $\vec p_3$ in the initial $\gamma p$ lab frame.

Assume now that the pion with momentum $\vec p_2$ is the one that decays into
$\gamma\gamma$. In the pion rest frame the two $\gamma$'s  will go back to back
and one $\gamma$ will have the momentum
\begin{eqnarray}
\vec{ \tilde p}_\gamma= \frac{m_{\pi^0}}{2}\left\{\begin{array}{c}
sin\theta_\gamma\,cos\varphi_\gamma\\
sin\theta_\gamma\,sin\varphi_\gamma\\
cos\theta_\gamma
\end{array}\right\},
\end{eqnarray}
with $\theta_\gamma,\,\varphi_\gamma$ angles of the photon in this one $\pi^0$
rest frame. Once again we boost this photon momentum to the frame where the pion
has momentum $\vec p_2
$\begin{eqnarray}
\vec p_\gamma=\left[\left(\frac{E_{2}}{m_{\pi^0}}-1\right)\frac{\vec{\tilde
p}_\gamma\cdot\vec p_2}{\vec p_2^2}+\frac{\tilde p_\gamma^0}{m_{\pi^0}}
\right]\vec p_2+\vec{\tilde p}_\gamma.
\end{eqnarray}

Since we can have a $\pi^0\gamma$ combination from either of the two $\pi^0$
or the two $\gamma$'s, we would obtain a combinatorial factor of four to account
for these possibilities.

All this said, the cross section for $\gamma p \to \pi^0\pi^0 p\to
\gamma\gamma\ \pi^0 p\to \gamma\pi^0+X$ reads 
\begin{eqnarray}
\sigma&=&4\,\frac{M^2_N}{s-M^2_N}\int\frac{d^3p_1}{(2\pi)^3}\frac{1}{2E( p_1)}
\left|T\right|^2\theta(M_{23}-2m_{\pi^0})  \\
&\times& \int_{-1}^1 d cos\theta \int_0^{2\pi}
d\varphi\, \tilde p_2 
\frac{1}{16\pi^2}\frac{1}{M_{23}}\frac{1}{4\pi}
\int_{-1}^1 d cos\theta_\gamma \int_0^{2\pi}
d\varphi_\gamma \nonumber
\label{eq:cros_sec_3}
\end{eqnarray}
recalling from Eq. (\ref{eq:invmass}) that in the lab frame
\begin{eqnarray}
M_{23}=s+M^2_N-2(E_{\gamma\, in}+M_N)E(p_1)+2\vec p_{\gamma\, in}\cdot\vec p_1
\label{eq:invmass_2}
\end{eqnarray}

Next one generates random numbers for $\vec p_1,\,\theta,\,\varphi,\,
\theta_\gamma,\,\varphi_\gamma$ with $|\vec p_1|$ restricted between zero and
\begin{eqnarray}
|\vec p_1|^{max}_{lab}=\frac{|\vec p_1|^{max}_{CM}+v\,E_{1\,CM}^{max}}{\sqrt{1-v^2}}
\end{eqnarray}
with $|\vec p_1|^{max}_{CM}$ and $E_{1\,CM}^{max}$ the maximum momentum and energy
allowed for the final proton in
the $\gamma p$ center of mass (CM) frame, corresponding to the case when the two
$\pi^0$ go together
\begin{eqnarray}
|\vec p_1|^{max}_{CM}=\frac{\lambda^{\frac12}(s,M^2_N,4m_{\pi^0}^2)}{2\sqrt s}
\end{eqnarray}
and $v$ is the velocity of the $\gamma p$ CM system measured in the lab frame
\begin{eqnarray}
v=\frac{|\vec p_{\gamma\,in}|}{E_{\gamma\,in}+M_N}
\end{eqnarray}

For each of these events we evaluate the invariant mass
\begin{eqnarray}
M_{inv}(\pi^0\gamma)=(p_\gamma+p_3)^2
\end{eqnarray}
and store the events, properly weighted by $|T|^2$ and phase space factors, in boxes of $M_{inv}(\pi^0\gamma)$ for a suitable
partition of $M_{inv}(\pi^0\gamma)$.
\subsection{The $\gamma p \to \pi^0\eta p$ reaction} 
We also evaluate the background for the $\gamma p \to \pi^0\eta p$ reaction.
In this case $\eta\to\gamma\gamma$ and we get one photon from there plus 
the $\pi^0$ to reconstruct the $\pi^0 \gamma$ invariant mass. The changes with respect to the
former reaction are minimal. Since now there is only one $\pi^0$, we do not have
to include the 1/2 symmetry factor and the
combinatorial factor of four before is now a factor of two and the mass of one
pion must be changed to the mass of the $\eta$ when needed. There are data for
this reaction in~\cite{mariana,Nakabayashi:2006ut,Ajaka:2008zz,ulrike}. There are also
recent detailed models for the reaction~\cite{Doring:2005bx,Doring:2006pt}
accounting fairly well for the cross section~\cite{mariana,Nakabayashi:2006ut} 
and the asymmetries~\cite{Ajaka:2008zz}. However, once again, for the present
problem it suffices to repeat the procedure done for the 
$\gamma p \to \pi^0\pi^0 p$ reaction in the former section taking a constant 
$|T|^2$ and implementing properly the phase space demanding that we reproduce
the data of~\cite{mariana,Nakabayashi:2006ut,ulrike}. 
\subsection{Background from extra sources}
As we shall see in the results section, the 
$\gamma p \to \pi^0\pi^0 p$ and $\gamma p \to \pi^0\eta p$ reactions can 
account for the background observed in the CBELSA/TAPS experiment~\cite{david,prl} 
in the region of the $\omega$ excitation and higher
 $\pi^0\gamma$ invariant masses. However, it does not account for the large
 background observed in the region of   $\pi^0\gamma$ invariant masses
 lower than $m_\omega$. Once again we resort to the findings of~\cite{david}
 suggesting that such events could come from reactions like $\gamma p\to
 \pi^0\pi^+ n$ with a misidentification of the neutron by a photon and other
 possible sources of $\gamma$ misidentification. In this part we do not make a
 theory since the events come from ignorance of the occurring reactions and
 misidentification of particles which have to do with the detector system.
 However, we would like to have $\pi^0\gamma$ events corresponding to this
 region in order to perform later on the mixed event analysis.

 To generate background events in this region we write the corresponding cross section as 
 \begin{eqnarray}
 \label{eq:sigma}
 \sigma&=&\int_{M_{inv}^{min}}^{M_{inv}^{max}} dM_{\pi\gamma} \
 |T(M_{\pi\gamma})|^2  \\
&\times& \int_{-1}^1d cos\theta
\int_0^{2\pi}d\varphi \int d^3P_{CM} \
\theta(500\,{\rm MeV/c}-|\vec P_{CM}|)\ ,
\nonumber
\end{eqnarray}
where $|T(M_{\pi\gamma})|^2$ is a function to be determined from experiment.
 $\theta,\,\varphi$ are the $\pi^0$ angles
in the $\pi^0\gamma$ rest frame. There the pion momentum is
\begin{eqnarray}
\vec{ \tilde p}_{\pi^0}= \tilde p_{\pi^0}\left\{\begin{array}{c}
sin\theta\,cos\varphi\\
sin\theta\,sin\varphi\\
cos\theta
\end{array}\right\},
\label{vectorform}
\end{eqnarray}
with
\begin{eqnarray}
\tilde p_{\pi^0}=\frac{M_{\pi\gamma}^2-m_{\pi^0}^2}{2M_{\pi\gamma}},~~~~~~~ 
\vec{\tilde p}_\gamma=-\vec{\tilde p}_{\pi^0}.
\end{eqnarray}
Eq.~(\ref{eq:sigma}) contains an integral over $\vec P_{CM}$ with a maximum of 
500\,MeV for $|\vec P_{CM}|$. This momentum represents the $\pi^0\gamma$ total
momentum in the $\gamma_{in}p$ CM frame.
 In this frame the momenta are more evenly distributed than in the lab frame and
then we take an isotropic distribution for  $\vec P_{CM}$ with the 
constraint $|\vec P_{CM}|<500$\,MeV/c. This is a conservative estimate that exceeds
the phase space of Eq.~(\ref{eq:sigma}) when $|\vec P|=|\vec p_{\pi^0}
+\vec p_\gamma|$ the total $\pi^0\gamma$ momentum in the $\gamma_{in} p$
lab frame is restricted to values smaller than 500 MeV/c.
Next we boost the $\pi^0$ and the $\gamma$ momenta from their CM frame to the 
$\gamma_{in} p$ CM frame where
 the $\pi^0\gamma$ system has momentum $\vec P_{CM}$. We have
\begin{eqnarray}
\vec p'_{\pi^0}&=&\left[\left(\frac{E_{\pi\gamma}}{M_{\pi\gamma}}-1\right)
\frac{\vec{\tilde
p}_{\pi^0}\cdot\vec P_{CM}}{\vec P^2_{CM}}+\frac{\tilde p_{\pi^0}^0}{M_{\pi\gamma}}
\right]\vec P_{CM}+\vec{\tilde p}_{\pi^0},\nonumber\\ \\
 E_{\pi\gamma}&=&\sqrt{\vec
P^2_{CM}+M^2_{\pi\gamma}}, \\
\vec p'_\gamma&=&\vec P_{CM}-\vec p_{\pi^0}.
\end{eqnarray}
The next step is the boost to the lab system where the $\gamma_{in}p$ has
momentum $\vec p_{\gamma\, in}$ and energy $E_{\gamma p}=p_{\gamma\,in}+M_N$,
hence
\begin{equation}
\vec p_{\pi^0}=\left[\left(\frac{E_{\gamma p}}{\sqrt s}-1\right)
\frac{\vec{
p}'_{\pi^0}\cdot\vec p_{\gamma\,in}}{\vec p^2_{\gamma\,in}}+\frac{
p_{\pi^0}^{\prime\,0}}
{\sqrt s}
\right]\vec p_{\gamma\,in}+\vec{ p}'_{\pi^0},
\end{equation}
and a similar one for $\vec p_\gamma$. On these $\gamma$ and $\pi^0$
momentum we enforce now the cut
\begin{equation}
\left|\vec p_{\pi^0}+\vec p_\gamma
\right|<500\,{\rm MeV/c}.
\end{equation} 
The function $T(M_{\pi\gamma})$ of Eq.~(\ref{eq:sigma}) 
is determined empirically such that the sum of
the cross section for $\gamma p\to \pi^0\pi^0 p$ plus $\gamma p\to \pi^0\eta p$,
plus the new one simulating $\gamma$ misidentification events, gives the total
experimental cross section of~\cite{david,prl}.

\begin{figure*}
\begin{center}
\resizebox{0.75\textwidth}{!}{%
  \includegraphics{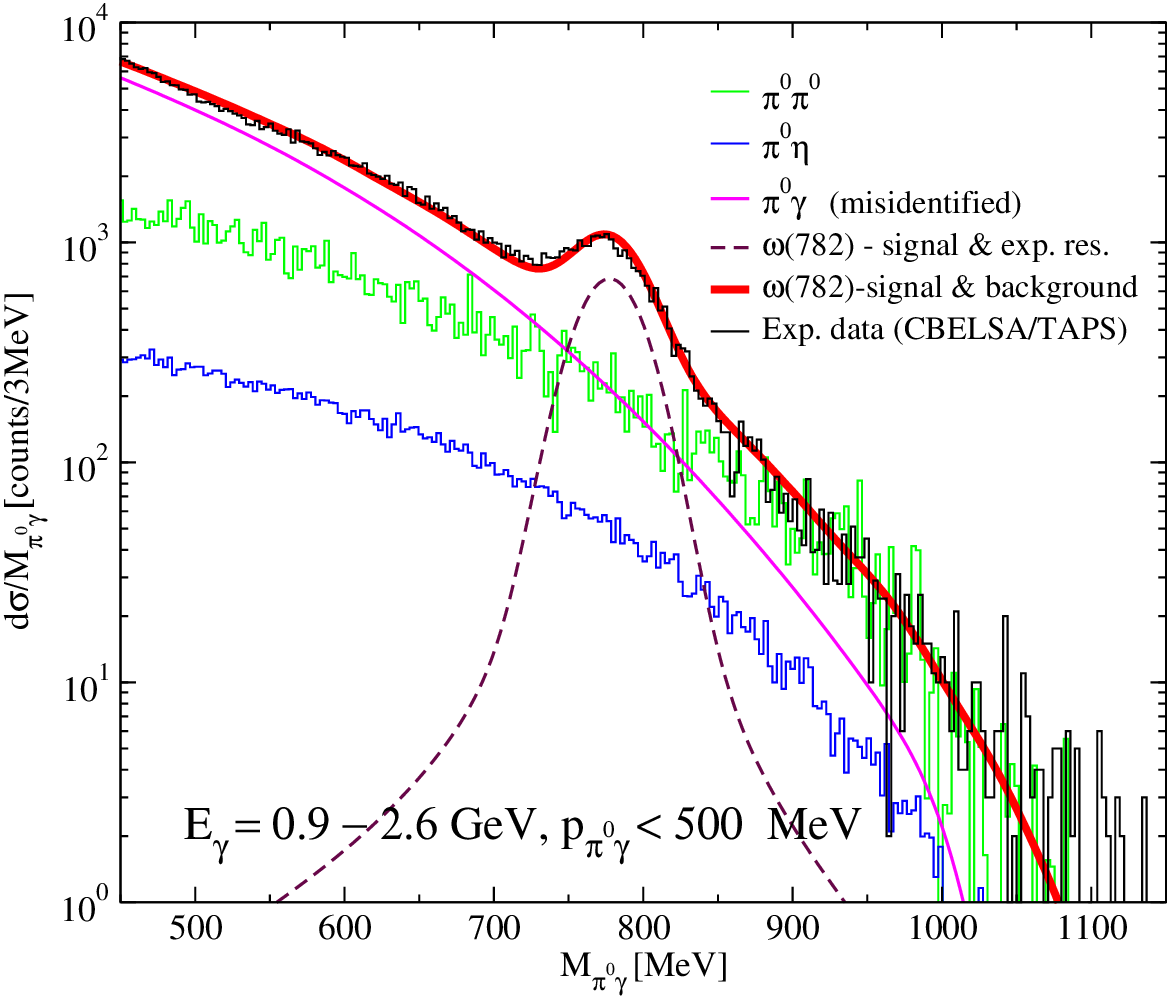}
}
\caption{The invariant mass
  $d\sigma/dM_{\pi^0\gamma}$ distribution in the reaction $\gamma p\to
  \pi^0\gamma p$. The signal and different background contributions are also shown.}
\label{fig:1}       
\end{center}
\end{figure*}
\section{Mixed events calculation}
In the mixed event simulation the idea is to obtain the background from the real
data by evaluating $M_{\pi^0\gamma}$ selecting the $\pi^0$ and the $\gamma$ 
from two different events. The invariant mass distribution is then given by
\begin{eqnarray}
M_{\pi^0\gamma}^2(ME)=(p_{\pi^0}(1)+p_\gamma(2))^2.
\label{eq:invmasspigamma}
\end{eqnarray}
There is abundant literature on the subject~
\cite{Drijard:1984pe,Fachini:2002iz,L'Hote:1992qz,vanEijndhoven:2001kx,Toia:2005vr} 
and it has become a popular instrument to determine background and 
isolate particular reactions that peak at a certain place.

In our case where all the integrals of the cross sections are performed by Monte
Carlo, the mixed event simulation is particularly simple to implement. The Monte
Carlo integrals are done by generating random events into a volume $V$ containing
the whole phase space and then the integral is given by the average value of the
integrand, that we shall denote as
$\widehat{|T|^2}$, times the volume
\begin{eqnarray}
\sigma=\frac{\sum_{i=1}^N \widehat{|T_i|^2}}{N}\,V,
\end{eqnarray}
where one understand that $\widehat{|T_i|^2}$ is zero if the event generated does not
belong to the phase space.
Assume now that we take pair of events $i,j$ corresponding to a reaction
channel. We have
\begin{eqnarray}
\sigma^2=\frac{\sum_{i=1}^N \widehat{|T_i|^2}}{N}\,V
\frac{\sum_{j=1}^N \widehat{|T_j|^2}}{N}\,V,
\end{eqnarray}
or equivalently
\begin{eqnarray}
\sigma=\frac{1}{\sigma}\,\sum_{i=1}^N\sum_{j=1}^N\frac{ \widehat{|T_i|^2}V
\widehat{|T_j|^2}V}{N^2}.
\label{eq:doblesigma}
\end{eqnarray}
We can then generate pairs of events in the phase space volume $V$ and the
former integral gives the cross section. Simultaneously with the evaluation
of the cross section we can obtain $M_{\pi^0\gamma}$ for each pairs of events
as done in Eq.~(\ref{eq:invmasspigamma}) and store the event, with its
corresponding weight, in a box of a certain $M_{\pi^0\gamma}$ 
value. After the double sum in Eq.~(\ref{eq:doblesigma}) we obtain
the normalized $d\sigma/dM_{\pi^0\gamma}$ distribution.

The generalization to four channels as we have in our case, $\gamma
p\to\pi^0\pi^0p$, $\gamma p\to\pi^0\eta p$, the channel from $\gamma$
misidentification and the $\gamma p\to \omega p\to\pi^0\gamma p$ channel is straightforward
\begin{eqnarray}
\sigma_{tot}=\frac{1}{\sigma_{tot}}\,\sum_\alpha\sum_\beta
\,\sum_{i=1}^N\sum_{j=1}^N\frac{ \widehat{|T_{i,\alpha}^{(1)}|^2}V
\widehat{|T_{j,\beta}^{(2)}|^2}V}{N^2},
\end{eqnarray}
where
\begin{eqnarray}
\sigma_{tot}=\sigma_1+\sigma_2+\sigma_3+\sigma_4,
\end{eqnarray}
and $\alpha,\beta$ run from 1 to 4. In order to obtain 
$d\sigma/dM_{\pi^0\gamma}$ from these mixed events we
evaluate again $M_{\pi^0\gamma}$ from Eq.~(\ref{eq:invmasspigamma}) and store the
events in  boxes of $M_{\pi^0\gamma}$ and we obtain at the end the histogram that
provides us with the normalized  $d\sigma/dM_{\pi^0\gamma}$ distribution from
mixed events.

\begin{figure*}
\begin{center}
\resizebox{0.75\textwidth}{!}{%
  \includegraphics{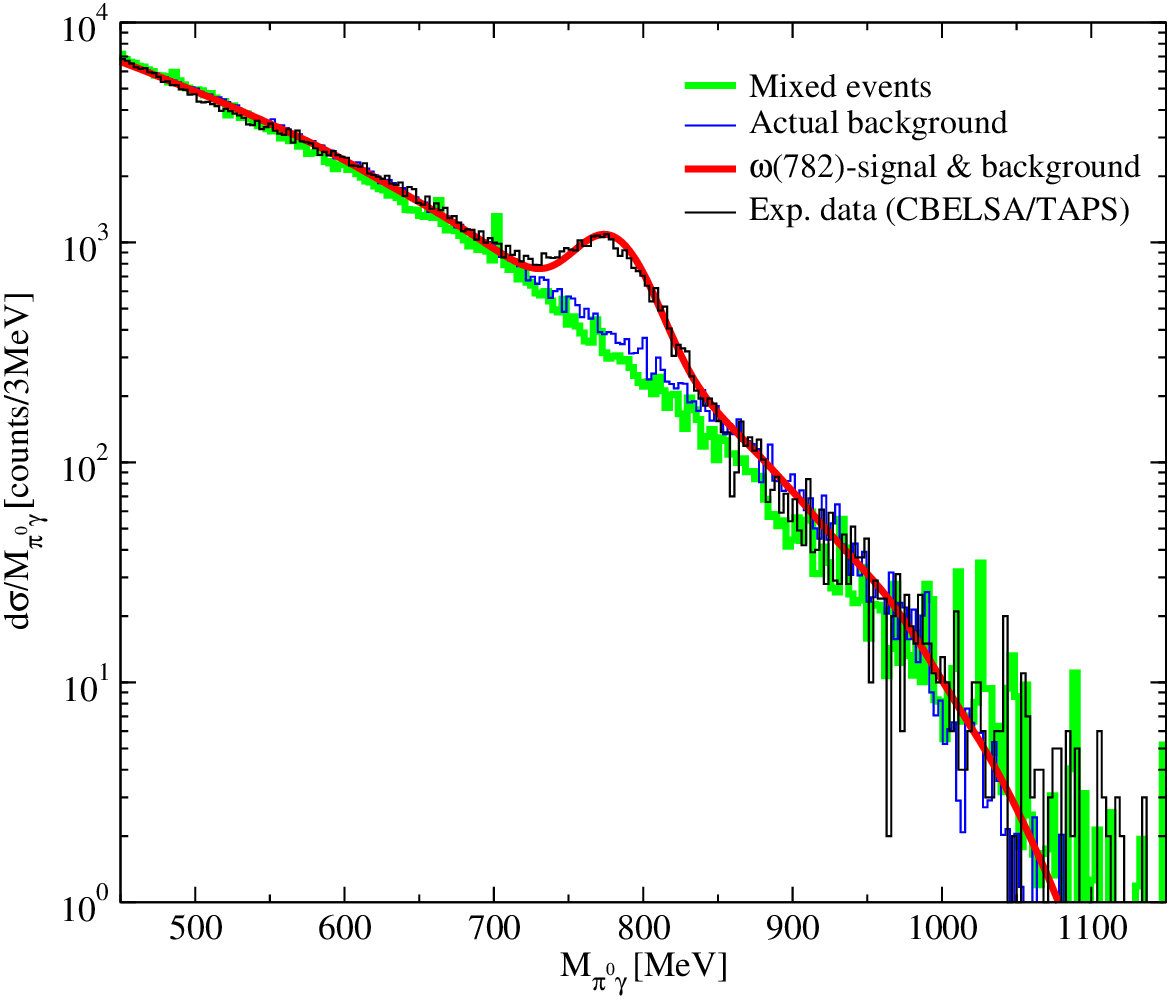}
}
\caption{The mixed event background with method I is shown, together with the actual background, the results of the model for $\omega$ signal plus background and the data.} 
\label{fig:2}       
\end{center}
\end{figure*}

\begin{figure*}
\begin{center}
\resizebox{0.75\textwidth}{!}{%
  \includegraphics{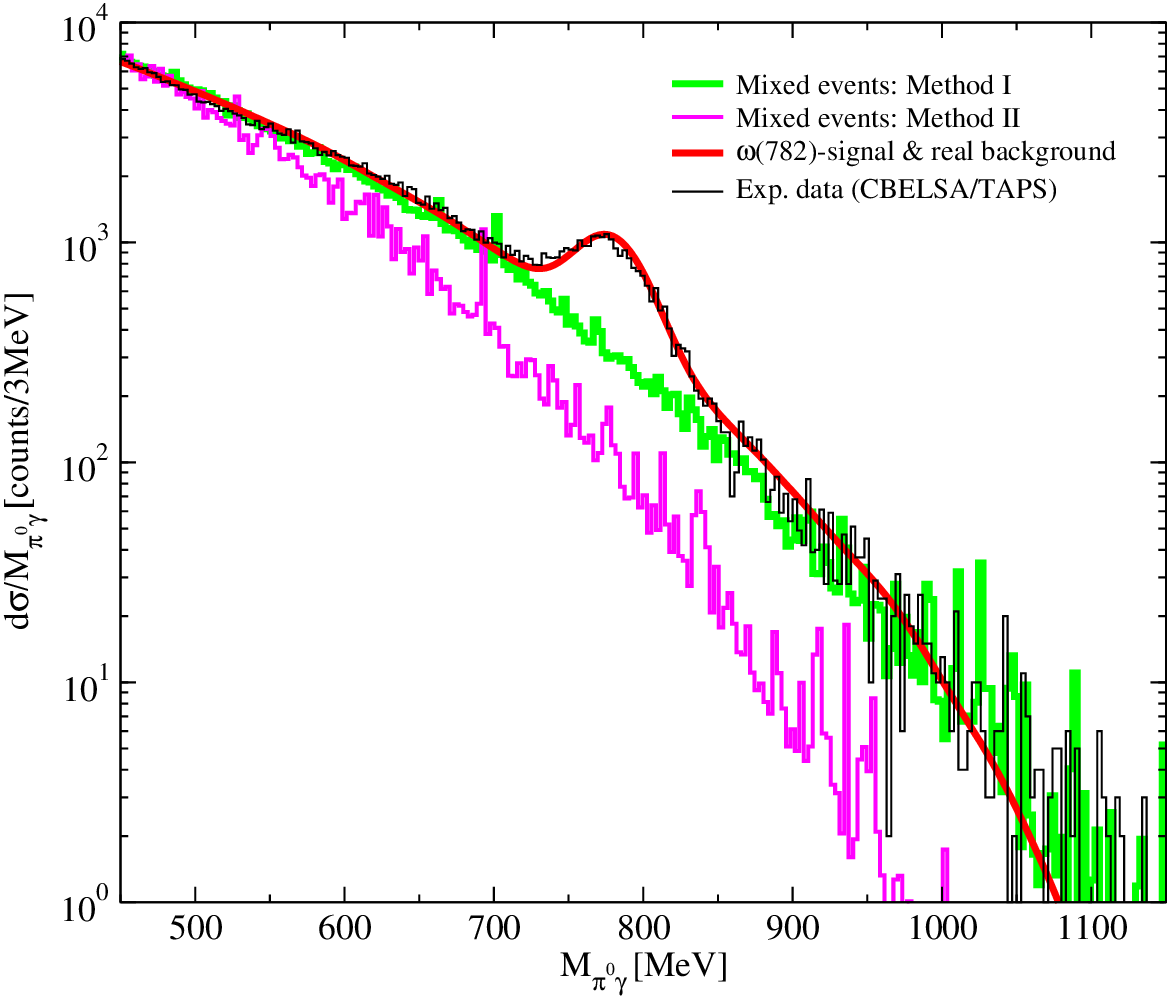}
}
\caption{The effect of the momentum cut on the mixed events before and after the
mixing (method II).}
\label{fig:3}       
\end{center}
\end{figure*}

\begin{figure*}
\begin{center}
\resizebox{0.75\textwidth}{!}{%
  \includegraphics{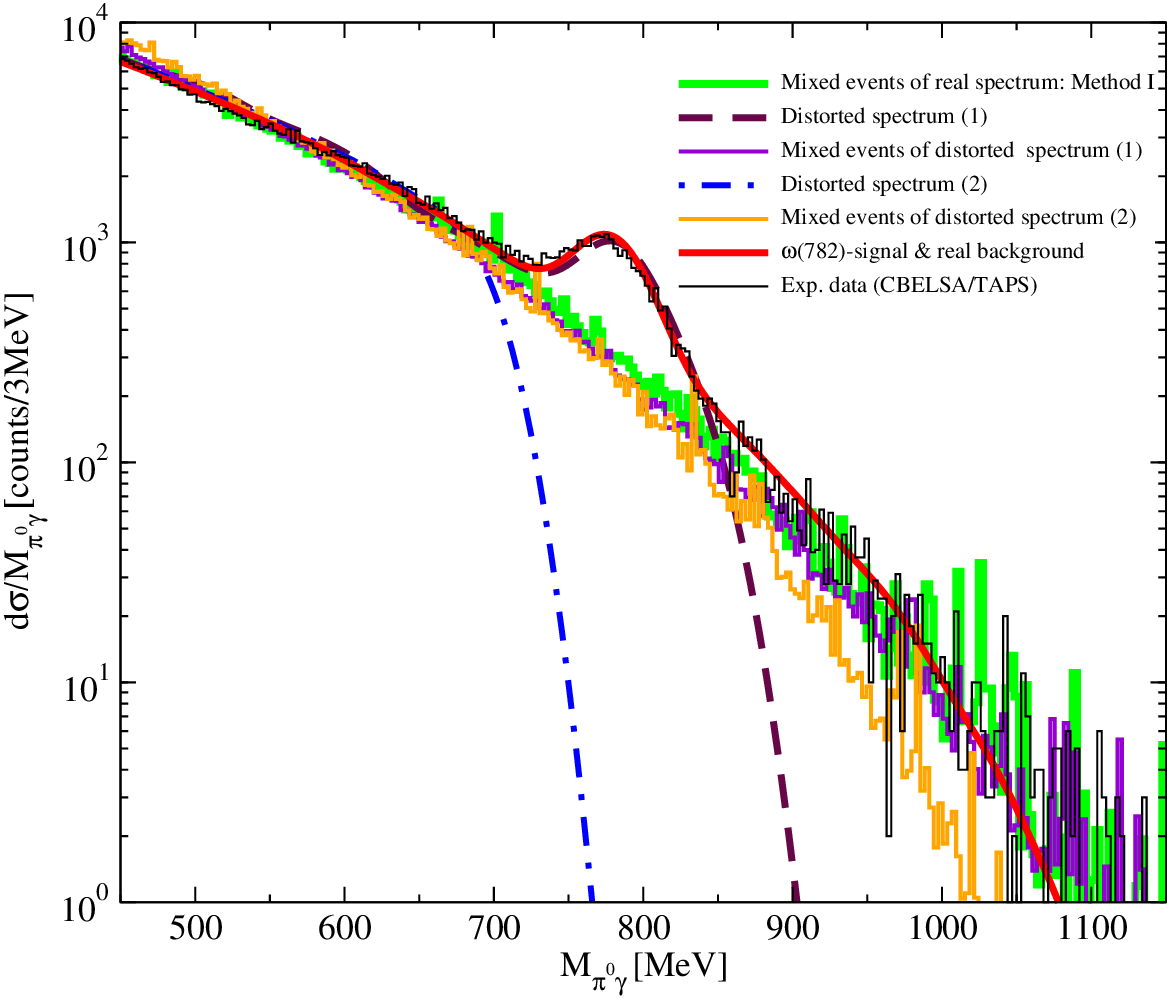}
}
\caption{The mixed event background with Method I for different input invariant mass distributions.}
\label{fig:4}       
\end{center}
\end{figure*}

\section{Results}
In Fig.~\ref{fig:1} the contributions from the $\omega$-signal and different
sources discussed above are compared with the experimental invariant mass
  $d\sigma/dM_{\pi^0\gamma}$ distribution in the reaction $\gamma p\to
  \pi^0\gamma p$ from CBELSA/TAPS experiment.
  
  The sources of background are $\gamma p\to \pi^0 \pi^0 p$ with either of the  two pions 
decaying into two $\gamma$, which was studied in section 2.1. This is the most 
important source of background in the region of the $\omega$ and beyond.  The other 
important source of background is the $\gamma$ misidentification studied in
Section 2.3.  This source competes with the former one in the region of the
omega and becomes dominant at smaller  $\gamma\pi^0$  invariant masses.
The third source considered is the one coming from $\gamma p\to \pi^0 \eta p$ 
followed by the $\eta$ decay into two $\gamma$. This source has a smaller
strength than the other two, but was found to be important to understand a
peak in the experiment at lower  $\gamma\pi^0$  invariant masses than the
omega in~\cite{Kaskulov:2006fi} when protons were measured in coincidence. The
omega signal comes from our study in~\cite{Kaskulov:2006zc}. The fit to the 
unnormalized data is done by adjusting the strength of the $\gamma p\to \pi^0
\pi^0 p$ source to the experiment distribution at large invariant mass, since 
this is the most important source in this invariant mass region. The source 
of the $\gamma p\to \pi^0 \eta p$, as well as the signal are rescaled
keeping their ratio, in order to keep the theoretical proportion between all these 
different sources. Finally, the source of misidentified $\gamma$ is added in 
order to complete a good description of the data.  As one can see from the 
figure, the agreement of the theoretical model with the experimental data 
is very good. Note that we also have adapted our theoretical set up to the 
experimental one by choosing $E_{\gamma}=0.9-2.6~ GeV$ and $p_{\pi^0 \gamma}=
|\vec p_{\pi^0}+
\vec p_\gamma| \leq 500 ~MeV $.

We should note that the experimental spectrum shown in Fig. \ref{fig:1} is not acceptance corrected. The reason is that experimentally one observes only three out of four photons in the final state due to the detector inefficiencies, or the overlap of photon clusters, and the latter depends on the energy of the photon \cite{marianahad}. Yet, in the unnormalized spectrum to which we make the fit, the differences are of the order of 20 \% from the lower mass part of the spectrum to the higher mass one, and have irrelevant consequences for the argumentations and conclusions that follow. The consideration of acceptance 
would be more important in case one would like to compare our different sources of background with the acceptance uncorrected experimental determinations \cite{marianahad}, which is not our concern here.

\section{Mixed events and different tests}
On the first hand we make the ME simulation that was done in~\cite{metagerice}
taking two independent events and demanding that 
\begin{eqnarray}
|\vec p_{\pi^0}^{\,(1)}+
\vec p_\gamma^{\,(2)}|<500\,{\rm MeV/c}
\label{eq:const}
\end{eqnarray}
for the mixed event. We call it method I. This choice has in principle a
conceptual flaw. Indeed, the curve in Fig.~\ref{fig:1}
corresponds to events in which one has imposed 
$|\vec p_{\pi^0}+\vec p_\gamma|<500$\,MeV/c in the $\pi^0$ and $\gamma$ momenta
of the same event. This restricts the phase space considerably. Now if one
imposes Eq.~(\ref{eq:const}) after the mixing, one can have both events (1) and (2)
or one of them that do not fulfill separately 
$|\vec p_{\pi^0}+\vec p_\gamma|<500$~MeV/c 
and, as a consequence, these are events which do not contribute  to the
distribution of Fig.~\ref{fig:1}. In other words, one can be using events that do
not provide any information to Fig.~\ref{fig:1} to obtain its corresponding
background through the mixed event method. Clearly in the extreme case that most
events in the ME simulation do not pass the individual
$|\vec p_{\pi^0}+\vec p_\gamma|<500$\,MeV/c test one would be obtaining the background of
the curve from a physical situation that has no relationship with the
distribution of Fig.~\ref{fig:1}. How far is one in practice from this
situation depends of course on the cut. 

In Fig.~\ref{fig:2} we show the results that we get from the mixed event 
simulation for the background, compared to the real background of the 
theoretical model. We can see that there is a remarkable agreement 
between the two in the whole range of invariant masses. We might conclude 
from there that the mixed event method is really good to reproduce the 
background.  Yet, let us investigate with more detail how this shape 
has been produced.  
Upon renormalization of the generated mixed event 
distribution we reproduce the real background, but we know that one is 
using for sure information not included in the spectrum of Fig.~\ref{fig:1}. 

Another way to proceed is to select two independent
events from Fig.~\ref{fig:1}, meaning that each one separately fulfills
$|\vec p_{\pi^0}+\vec p_\gamma|<500$\,MeV/c, and then reconstruct the invariant mass
of Eq.~(\ref{eq:invmasspigamma}) for the mixed events, imposing also the cut of
Eq.~(\ref{eq:const}) to the pair of events of the mixing. We call that  
method II. This would correspond to a ME reconstruction from experimental 
events that have been filtered with the $|\vec p_{\pi^0}+\vec p_\gamma|<500$\,
MeV/c condition, which imposes a certain correlation, which might be
undesired, in the events chosen for the mixing. The results can be seen in
Fig.~\ref{fig:3}.  Now we normalize 
the background at low invariant masses where it is maximum. Then we observe 
that in the rest of the invariant mass region there is a clear disagreement 
of this new mixed events background with the real one. Hence, the result has 
been a very poor reproduction of the real background by the mixed event method
II.

 Going back to method I, and in order to understand what is really happening,
we have conducted another test. Let us 
realize that the mass distribution of 
Fig.~\ref{fig:1}  is exponential and there are three orders of magnitude
difference between the strength of $d\sigma/dM_{\pi^0\gamma}$ at low and large
invariant masses. From pure statistics it looks quite logical that if we take
two independent events to reconstruct the mixed event $\pi^0\gamma$ invariant
mass, these two events belong to the region of the spectrum that has larger
cross section, even if the mass that we obtain corresponds to the large 
invariant mass where  $d\sigma/dM_{\pi^0\gamma}$ is small. In other words, it is
perfectly acceptable that the background that one obtains in the large invariant
mass region is largely determined by events far away from this region, sitting
at much lower individual  $\pi^0\gamma$ invariant
masses. If this were the case  one would be attributing the background in the
high invariant mass region to different reactions than those responsible for it
and hence one would be distorting the physics of the process. Certainly in such a
 case there would be an interesting side effect: the distribution obtained with
 the mixed event method at large invariant masses would be largely insensitive
 to the actual background contributing in that region. In this case the ME
 method would thus render a background in this region that has nothing to do
 with the actual one.

 In order to illustrate more dramatically the problem, we change arbitrarily the
 background of our model at large $M_{\pi^0\gamma}$ by imposing 
 \begin{equation}
 |T|^2\to |T|^2\ f(M_{\pi^0\gamma})
 \end{equation}
where $f(M_{\pi^0\gamma})$ is a distortion factor. We consider two sharp cuts with
\begin{eqnarray}
 (1)~~f(M_{\pi^0\gamma})=\left\{\begin{array}{ll}
 1 &{\rm\ \ \  for}\ M_{\pi^0\gamma}<850\,{\rm MeV}\\
 0&{\rm\ \ \  for}\ M_{\pi^0\gamma}>850\,{\rm MeV}
 \end{array}\right.
 \label{eq:chback}
 \end{eqnarray}
and 
\begin{eqnarray}
 (2)~~f(M_{\pi^0\gamma})=\left\{\begin{array}{ll}
 1 &{\rm\ \ \  for}\ M_{\pi^0\gamma}<750\,{\rm MeV}\\
 0&{\rm\ \ \  for}\ M_{\pi^0\gamma}>750\,{\rm MeV}
 \end{array}\right.
 \label{eq:chback2}
 \end{eqnarray}
 The distortion factor in Eq.~(\ref{eq:chback}) cuts off the background at
 higher invariant masses beyond the $\omega$-signal and (\ref{eq:chback2})
 removes both the signal and the background.

 In Fig.~\ref{fig:4} we show the results for the background from the ME
 method, with method I and the real data, compared with those obtained 
with the distorted spectra of Eqs.~(\ref{eq:chback})
and~(\ref{eq:chback2}). Note that the sharp cut off becomes in the figure 
a smoother fall down because we implement the folding of the $\pi^0 \gamma$ 
invariant mass with the experimental resolution of \cite{prl} of $50 ~MeV$. 
What we see in Fig. \ref{fig:4} is that the ME method output barely
 changes with the actual background from
 Eqs.~(\ref{eq:chback},\ref{eq:chback2}), or in other words,
 that the ME method is unable to produce the actual background. It produces a
 background largely tied to the events at low invariant mass and is unsuited to
 produce a realistic background in the region of large $M_{\pi^0\gamma}$ masses.
   
In order to test the former suggestion that the background at large invariant
 masses in the ME method is tied to events at low invariant masses, 
we construct the correlation matrix $C(M_{\pi\gamma}(ini), M_{\pi\gamma}(fin))$ where $M_{\pi\gamma}(ini)$
 refers to any of the two events used in the mixed event simulation and 
 $M_{\pi\gamma}(fin)$ refers
 to the ME final invariant mass determined through Eq.~(\ref{eq:invmasspigamma}).
 
\begin{figure*}
\begin{center}
\resizebox{0.55\textwidth}{!}{%
  \includegraphics{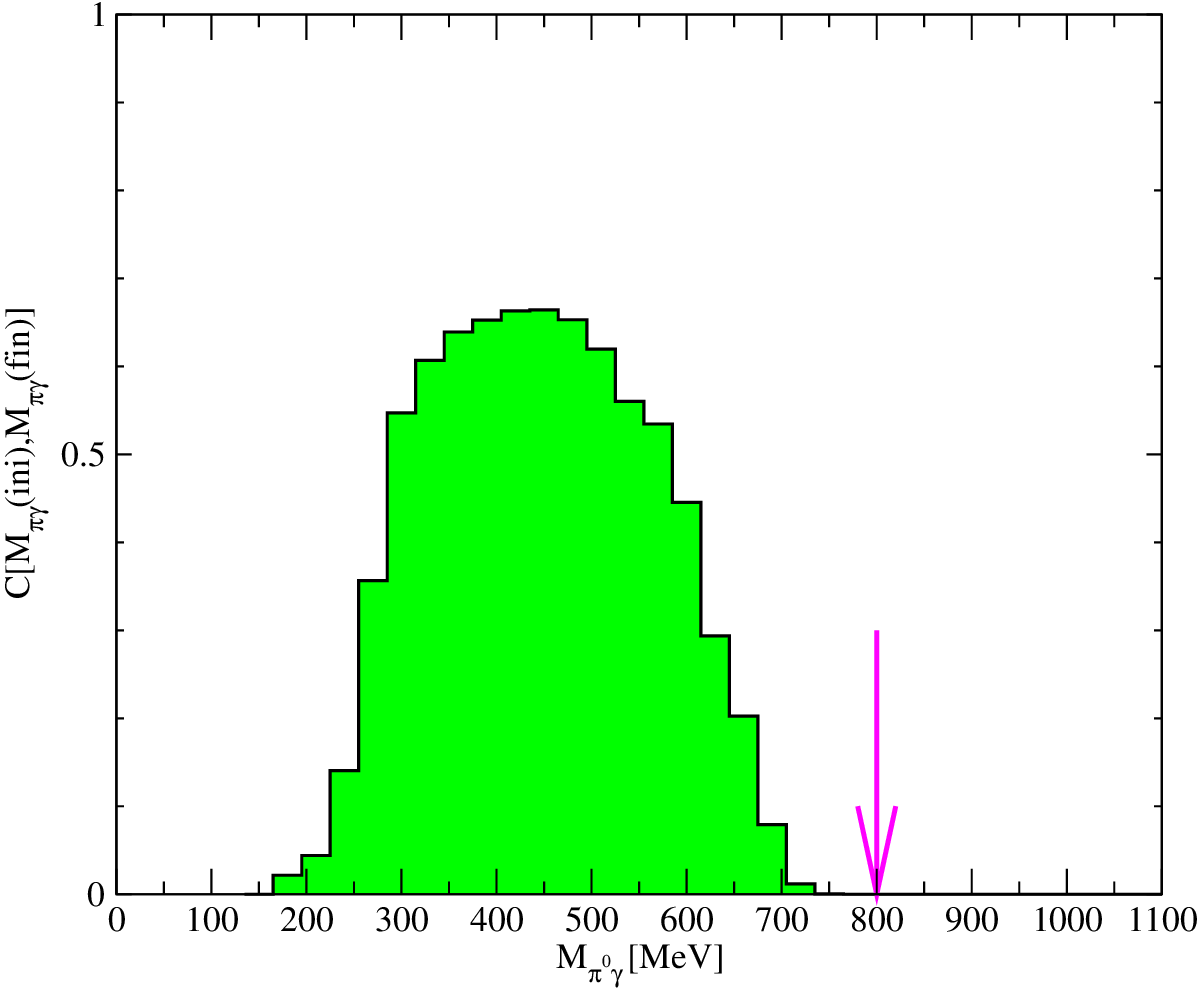}
}
\caption{The correlation function $C(M_{\pi\gamma}(ini), M_{\pi\gamma}(fin))$
  (see the details in the text)} 
\label{fig:corr}       
\end{center}
\end{figure*}

 In Fig.~\ref{fig:corr} we plot the correlation function. We have taken
 $M_{\pi \gamma}(fin)= 800~MeV$.  
Then keeping this variable fixed we plot on the y-axis in an arbitrary scale
the 
number of events (summing the two events used in the mixing) that would have a 
certain   $M_{\pi \gamma}(ini)$. As we can see, the initial events used in the mixing accumulate in 
the region where  $M_{\pi \gamma}(ini)$ is about 400-500 MeV.  This 
certainly is distorting the physics of the problem, since the background 
associated to the region $M_{\pi \gamma}(fin)= 800~MeV$ is generated after 
mixing by events around 400 MeV, where the origin of the background is 
quite different from the real one around 
$M_{\pi \gamma}(fin)= 800~MeV$.

\section{Summary}
    In summary, what we have seen is that due to the peculiar shape of the
    background 
in the present process and the fast drop as a function of 
$M_{\pi \gamma}$, the mixed event method is unsuited to provide an even
qualitative reproduction of the real background of the process.  Even if a first run
seemed encouraging because it gave a good reproduction of the background, 
further insight into the method revealed its flaws since we could prove that 
different methods to do the cuts gave rise to very different mixed events 
background. Further we could prove that the results provided 
by the mixed event method were practically insensitive to the value of the 
real background beyond 750 MeV, to the point that we could take any arbitrary 
background as input in that region and the mixed event method would always 
provide the same background, with no resemblance to the one that it was 
supposed to reproduce. The study of the correlations of events gave us an 
explanation for this finding, since we saw that even at large invariant
masses, the events of the mixing that generated the final  $M_{\pi \gamma}$ 
were  collected for the region of $M_{\pi \gamma}(ini)$ around 400-500 MeV.  
In that region the origin of the background is very different from the one for 
the real events at large invariant masses, such that the mixing event method 
not only produces an unrealistic numerical background, but gets it from 
physical processes quite different from those responsible for the real 
background at large invariant masses, thus grossly distorting the physics of the process.

%
%
%
%
%
%
%
%
%
%




\begin{center}
{\bf Acknowledgments} 
\end{center}
We would like to thank V. Metag, M. Nanova, S. Friedrich and M. Kotulla for discussions on the issue and
on the experimental data. Useful information on mixed events was provided 
by C. Djalali and R. Nasseripour.
This work was supported by DFG through the SFB/TR16, by DGI and FEDER funds, under contracts
 FIS2006-03438, FPA2007-65748, The Generalitat Valenciana in the Prometeo Program and the Spanish Consolider-Ingenio 2010
Programme CPAN (CSD2007-00042), by  Junta de Castilla y Le\'on under 
contract SA 016A07 and GR12, and it is part of the EU integrated infrastructure
initiative Hadron Physics Project under contract number
RII3-CT-2004-506078.   



%
%

\end{document}